\documentclass[
showpacs,
floatfix,
aps,
prl,
amsmath,
twocolumn,
superscriptaddress,
amssymb,
]{revtex4-1}
\usepackage{amssymb}
\usepackage{latexsym}
\usepackage[dvips]{graphicx}
\usepackage{amsmath}
\usepackage{graphicx}
\usepackage{dcolumn}
\usepackage{amsfonts}
\usepackage{bm}
\usepackage{epsfig}

\newcommand{\be}{\begin{equation}}
\newcommand{\ee}{\end{equation}}
\newcommand{\bea}{\begin{eqnarray}}
\newcommand{\eea}{\end{eqnarray}}

\def\eac{\epsilon_\omega}
\def\edc{\epsilon_j}

\def\mus{\mu_S}
\def\mul{\mu_L}

\def\oc{\omega_{\mbox{\scriptsize {c}}}}

\def\rc{R_{\mbox{\scriptsize {c}}}}

\def\tq{\tau_{\mbox{\scriptsize {q}}}}

\newcommand{\req}[1]{Eq.\,(\ref{#1})}

\newcommand{\rfig}[1]{Fig.\,\ref{#1}}

\newcommand{\rref}[1]{Ref.\,\onlinecite{#1}}

\begin{document}

\title{Nonlinear transport in 2D electron gas exhibiting colossal negative magnetoresistance}

\author{Q.~Shi}
\affiliation{School of Physics and Astronomy, University of Minnesota, Minneapolis, Minnesota 55455, USA}
\author{M.~A.~Zudov}
\affiliation{School of Physics and Astronomy, University of Minnesota, Minneapolis, Minnesota 55455, USA}
\author{L.~N.~Pfeiffer}
\affiliation{Department of Electrical Engineering, Princeton University, Princeton, New Jersey 08544, USA}
\author{K.~W.~West}
\affiliation{Department of Electrical Engineering, Princeton University, Princeton, New Jersey 08544, USA}

\begin{abstract}
We report on nonlinear transport measurements in a GaAs/AlGaAs quantum well exhibiting a colossal negative magnetoresistance effect. 
Under applied dc bias, the magnetoresistance becomes nonmonotonic, exhibiting distinct extrema that move to higher magnetic fields with increasing current.
In the range of magnetic fields corresponding to the resistivity minimum at zero bias, the resistivity increases linearly with current and  the rate of this increase scales with the inverse magnetic field.
The latter observation is consistent with the theory, proposed more than 35 years ago, considering classical memory effects in the presence of strong, dilute scatterers
\end{abstract}

\received{September 19, 2014}

\maketitle

The interest to low-field magnetotransport in two-dimensional electron systems (2DES) has been recently revived owing to several experiments reporting unexpectedly strong negative magnetoresistance in GaAs/AlGaAs heterostructures \citep{dai:2010,hatke:2011b,dai:2011,bockhorn:2011,hatke:2012a,mani:2013,shi:2014a,bockhorn:2014}.
One prominent example is the observation of a colossal negative magnetoresistance (CNMR), which is marked by a sharp drop of the resistivity $\rho(B)$ followed by a saturation at the magnetic field $B \approx B_{\star} \approx 1$ kG, close to $\rho_{\star} \equiv \rho(B_{\star}) \lesssim 0.1 \rho_0$, at temperature $T \lesssim 1$ K \citep{shi:2014a}.
While classical memory effects due to a unique disorder landscape appear to be the most likely origin of the observed CNMR, comparison with existing theories \citep{baskin:1978,mirlin:2001,polyakov:2001} revealed huge discrepancy in the characteristic $B$ range where the effect is expected to occur.
It is thus clear that more studies are needed to shed light on this mysterious phenomenon.
In particular, it is very desirable to get insight into the specifics of the underlying disorder potential in the 2DES exhibiting CNMR.

The most frequently used and readily available characteristic of the disorder is the electron mobility $\mu =(en_e\rho_0)^{-1}$, where $n_e$ is the electron density and $\rho_0$ is the resistivity at $B=0$.
At low $T$, the mobility can be expressed as $\mu^{-1} = \mul^{-1} +\mus^{-1}$, where $\mul$ and $\mus$ account for scattering off \emph{long-range} (smooth) disorder, e.g. from the remote ionized impurities, and \emph{short-range} (sharp) disorder, e.g. from the residual background impurities, respectively.
Since magnetoresistance sensitively depends on the interplay between smooth and sharp contributions \citep{baskin:1978,mirlin:2001}, it is important to know not only the total $\mu$, but also at least one of its constituents, i.e. $\mul$ or $\mus$.

In principle, $\mus$ can be obtained from non-linear transport measurements, which are known to reveal Hall-field induced resistance oscillations (HIRO) \citep{yang:2002,bykov:2005c,zhang:2007a,vavilov:2007,hatke:2009c,hatke:2010a,hatke:2011a,dmitriev:2012}. 
HIRO appear in differential resistivity $r$ and originate from electron transitions between Hall field-tilted Laudau levels due to electron backscattering off impurities.
The corresponding scattering rate relies on the commensurability between the cyclotron diameter $2\rc$ and the spatial separation between the levels, and, as a result, is a periodic function of $\edc \equiv 2 e E \rc/\oc$, where $E$ is the Hall field and $\oc$ is the cyclotron frequency. 
Since backscattering is strongly dominated by sharp disorder, the HIRO amplitude is proportional to $\mus^{-1}$.
The full result reads \citep{vavilov:2007}:
\be
\frac r \rho_0 = \frac {(4 \lambda)^2} \pi \frac \mu {\mus} \cos 2\pi\edc\,,
\label{eq.hiro}
\ee
where $\lambda = \exp(-\pi/\oc\tq)$ is the Dingle factor and $\tq$ is the quantum lifetime.
The analysis of the HIRO amplitude can therefore be used to obtain both $\tq$ and $\mus$, which are essential in understanding the correlation properties of the disorder potential.

In this Rapid Communication we report on nonlinear transport measurements in a Hall bar-shaped 2DES exhibiting CNMR, marked by $\rho_{\star} \sim 0.1 \rho_0$ at $B_\star \simeq 1$ kG \citep{shi:2014a}, over a wide range of $B$ and direct currents $I$.
While the differential resistivity exhibits at least two distinct types of extrema, which both move to higher $B$ with increasing $I$, none of them can be described by \req{eq.hiro}.
On the other hand, in a wide range of $B$, corresponding to the broad resistivity minimum at zero bias, $\rho$ increases linearly with $I$ and the rate of this increase scales with $1/B$.
This remarkable finding is in excellent agreement with the theory considering classical memory effects in the presence of strong, large, and dilute scatterers, which was put forward more than 35 years ago \citep{baskin:1978}.

Our sample is a Hall bar (width $w=200$ $\mu$m) fabricated from a symmetrically doped, 29 nm-wide GaAs/ AlGaAs quantum well, with the Si $\delta$-doping layers separated from the 2D channel by $d = 80$ nm.
At $T \simeq 1.5$ K, our 2DES has $\mu\approx 1.1 \times10^{6}$ cm$^2$/Vs and density $n_e \approx 3.0 \times10^{11}$ cm$^{-2}$.
Measurements of the differential resistivity $r = dV/dI$ were performed in sweeping $B$ using a standard lock-in technique at $T \approx 1.5$ K at $I$ up to 350 $\mu$A.
Photoresistance was measured under continuous illumination by microwaves of frequency $f = 86$ GHz.

\begin{figure}[t]
\includegraphics{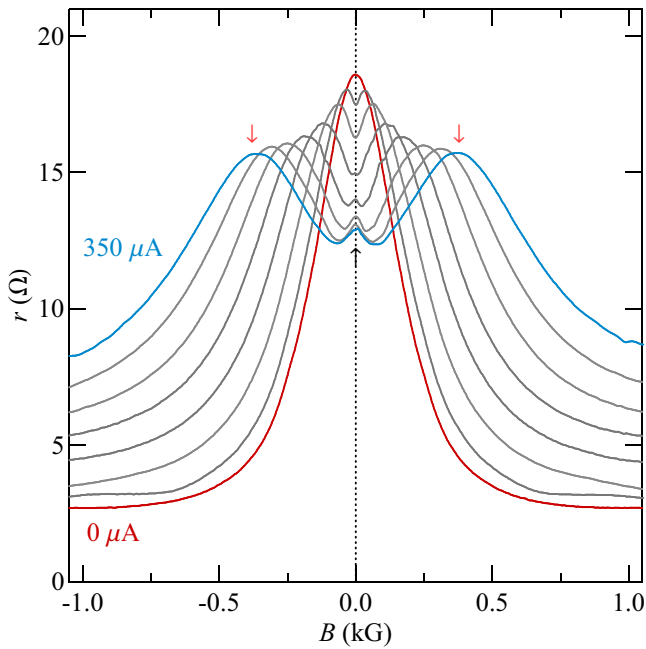}
\vspace{-0.05 in}
\caption{(Color online) 
Differential resistivity $r(B)$ at different $I$ from 0 to 350 $\mu$A, in steps of 50 $\mu$A, at $T = 1.5$ K.
}
\vspace{-0.15 in}
\label{fig1}
\end{figure}
In \rfig{fig1} we present the differential resistivity $r(B)$ measured at $T = 1.5$ K under different $I$ from $0$ to 350 $\mu$A, in steps of 50 $\mu$A. 
At zero current, $r(B) \equiv \rho(B)$ exhibits CNMR \citep{shi:2014a}, which is marked by a sharp drop of the resistivity terminating at $\rho_{\star} \equiv \rho(B_{\star}) \approx 0.1 \rho_0$, where $B_\star \approx 1$ kG. 
Under applied current, a local minimum develops in $r(B)$ at $B = 0$ as the ``zero-$B$ peak'' splits into two. 
With increasing $I$, the peaks move to higher $B$ (cf. $\downarrow$) and  the minimum at $B=0$ evolves back to a local maximum (cf. $\uparrow$).
This behavior is somewhat reminiscent to the splitting of the ``zero-$B$ peak'' observed at $2 {\rm~K} \lesssim T \lesssim 15 {\rm~K}$ at zero dc bias \citep{shi:2014a}. 

\begin{figure}[t]
\includegraphics{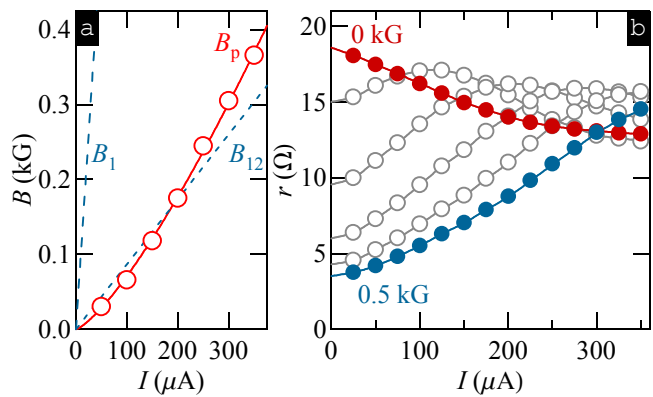}
\vspace{-0.15 in}
\caption{(Color online) 
(a) Magnetic fields corresponding to the dc-induced peaks (cf. $\downarrow$ in \rfig{fig1}) $B_{\rm p}$ (open circles), to the first, $B_1$, and twelfth, $B_{12}$, HIRO maxima (dashed lines) as a function of $I$. 
The fit to $B_{\rm p} \sim I^\alpha$ (solid line) yields $\alpha \approx 1.3$.
(b) $r(I)$ at different $B$, from 0 to 0.5 kG, in steps of 0.1 kG.
}
\vspace{-0.15 in}
\label{fig2}
\end{figure}

The observed dc-induced peaks cannot be attributed to HIRO as they occur at $B$ which is an order of magnitude lower than $B$ at which the fundamental ($\edc = 1$) HIRO peak is expected to occur.
To illustrate this point, we present in \rfig{fig2}(a) the magnetic field $B_{\rm p}$ at which the peak occurs (open circles) as a function of $I$. 
For comparison, we also include the expected magnetic fields for the fundamental and the twelfth HIRO peaks (dashed lines, marked by $B_1$ and $B_{12}$, respectively), which were calculated, per \req{eq.hiro}, using $B_i = B_1/ i$, $B_1 = 2(m^\star/e^2)\sqrt{2\pi/n_e}(I/w)$ \citep{zhang:2007a,vavilov:2007}.
Indeed, $B_{\rm p} \ll B_1$ and the peak is better described by $\edc \approx 11-13$. 
However, the dependence of $B_{\rm p}$ on $I$ is not linear and the power-law fit (solid line), $B_{\rm p} \sim I^\alpha$, yields $\alpha \approx 1.3$. 
We thus conclude that the observed peak differ from HIRO both quantitatively and qualitatively.

An alternative way to examine the evolution of $r(B)$ with current is presented in \rfig{fig2}(b) which shows $r(I)$ at different $B$ from 0 to 0.5 kG, in steps of 0.1 kG.
As we have already observed in \rfig{fig1}, $r_0 \equiv r (B = 0)$ monotonically decreases with current and, eventually, tends to saturate. 
This behavior is in contrast to the usually observed increase of $r_0(I)$, which can be expected to occur due to Joule heating \citep{zhang:2007a}. 
However, at finite $B$, $r(I)$ initially increases and develops a maximum which moves to higher $I$ with increasing $B$.
This maximum is just another manifestation of the peak shown in \rfig{fig1}.

Another characteristic feature of the data in \rfig{fig1} and \rfig{fig2}(b) is the absence of HIRO, which are usually seen in this range of $B$ and $I$ \citep{zhang:2007a}.
In particular, at $B = 0.5$ kG HIRO would manifest as $I$-periodic oscillations with the maxima appearing at integer multiples of $\approx 50$ $\mu$A and the minima in between.
We, however, find that $r(I)$ monotonically increases with no signature of oscillations.

\begin{figure}[t]
\includegraphics{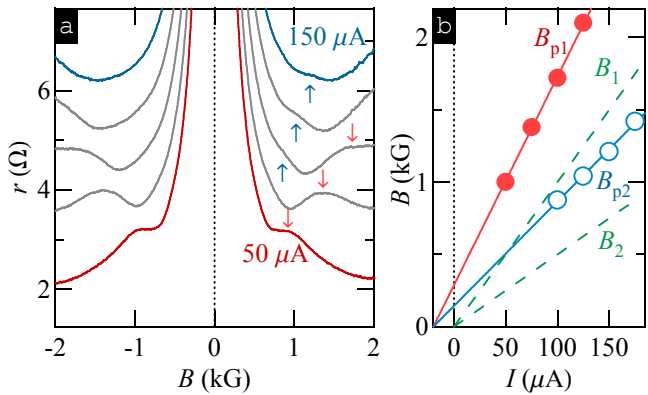}
\vspace{-0.15 in}
\caption{(Color online) 
(a) Differential resistivity $r(B)$ at different $I$ from 50 to 150 $\mu$A, in steps of 25 $\mu$A, at $T = 1.5$ K.
(a) Magnetic fields corresponding to the dc-induced peaks $B_{\rm p1}$ (solid circles) [cf. $\downarrow$ in \rfig{fig3}(a)], $B_{\rm p2}$ (open circles) [cf. $\uparrow$ in \rfig{fig3}(a)], and to the first, $B_1$, and the second, $B_2$, HIRO maxima, (dashed lines) as a function of $I$. 
Solid lines are linear fits to the data.
}
\vspace{-0.15 in}
\label{fig3}
\end{figure}

At higher $B$, some dc-induced oscillatory features do appear in $r(B)$, but these features cannot be attributed to conventional HIRO either.
In \rfig{fig3}(a) we present $r(B)$ at different $I$ from 50 to 150 $\mu$A, in steps of 25 $\mu$A, measured at $B$ up to $2$ kG.
These data clearly show another dc-induced peak (cf. $\downarrow$) which moves to higher $B$ with increasing $I$.  
Closer examination of the data reveals another, weaker maximum (cf. $\uparrow$),  emerging at lower $B$.
To see if these two peaks originate from HIRO, we construct \rfig{fig3}(b) showing their respective magnetic fields, $B_{\rm p1}$ (solid circles)and $B_{\rm p2}$ (open circles), as a function of $I$.
Even though both peaks exhibit linear dependencies on $I$, with the slopes differing by about a factor of two, they cannot be attributed to regular HIRO for the following reasons.
First, linear fits (solid lines) \emph{do not} extrapolate to the origin.
Second, the slopes of both fits are noticeably higher than what is expected from HIRO.
Indeed, positions of the HIRO maxima in our 2DES are expected to follow $B_i = \beta I/i$, with $\beta \approx 10$ G/$\mu$A.
The observed peaks, on the other hand, follow $B_{\rm pi} \approx \beta (I+I_0)/i$, where $\beta \approx 14.4$ G/$\mu$A and $I_0 \approx 20$ $\mu$A.

\begin{figure}[t]
\includegraphics{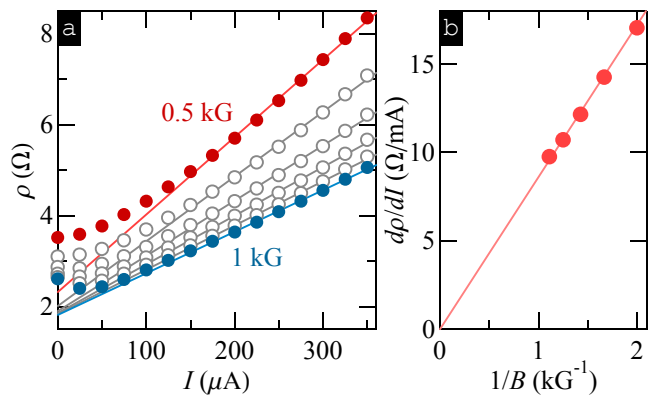}
\vspace{-0.05 in}
\caption{(Color online) (a) $\rho(I)$ (circles) at different $B$, from 0.5 kG to 1 kG, in steps of 0.1 kG.
(b) The slope of $\rho(I)$, obtained from linear fits (solid lines)  versus $1/B$.
Solid line is the linear fit, $d\rho/dI = \beta/B$, where $\beta \approx$ 0.94 $\Omega$ T
/mA.
}
\vspace{-0.15 in}
\label{fig4}
\end{figure}

Motivated by monotonic, nearly linear increase of $r(I)$ at $B = 0.5$ kG, we examine $r(I)$ in more detail at various $B$ within the broad minimum of $\rho(B)$ observed at $I=0$.
To facilitate the comparison with the theoretical model, discussed below, we convert $r(I)$ to $\rho(I) = I^{-1}\int_0^I r(I')dI'$ and present the result (circles) in \rfig{fig4}(a) at different $B$ from 0.5 to 1 kG, in steps of 0.1 kG.
We find that the data at all $B$ are well described by linear dependencies, starting from some offset current which drops from $\approx 200$ $\mu$A at $B = 0.5$ kG to $\approx 100$ $\mu$A at $B = 1$ kG.
We also observe that the slope of these dependencies monotonically decreases with $B$.
To examine how the slope depends on $B$, we fit the linear portions of the data (cf. straight lines) and then plot the obtained first derivative $d\rho/dI$ as a function of $1/B$ in \rfig{fig4}(b).
Since all of the data points fall on a straight line extrapolating to the origin, we conclude that $d\rho/dI \sim 1/B$. 

As we show below, this result agrees with the theory of negative magnetoresistance considering classical localization in the presence of strong, dilute scatterers (radius $a$, 2D density $n_i \ll a^{-2}$) \cite{baskin:1978,dmitriev:2001,dmitriev:2002}.
According to this model, negative magnetoresistance occurs because the probability for an electron to return to the same impurity increases with $B$ and, as a result, the probability for an electron to scatter off other impurities is reduced. 

Upon application of electric and magnetic fields, the electron motion is described by $\bf{E} \times \bf{B}$ drift; the electron moves along a helical trajectory with a pitch $\delta = 2\pi v_d/\oc$, where $v_d = E/B$ is the drift velocity of the electron guiding center.
If $\delta \ll a$, an electron returns to the same impurity and remains localized. 
At higher dc bias, but still such that $\delta < 2 \pi a$, an electron will leave an impurity after a short time ($\sim 2a/v_d$) and continue drifting until it collides with another impurity. 
The average time between collisions can then be estimated as $\tau_{E} = l/v_d$, where $l \approx (2\rc n_i)^{-1}$ is the mean free path of the electron guiding center.
After the collision, the guiding center is shifted by $\sim$$\rc$ and the drift continues till the next collision.  
Under these conditions, $\tau_E$ plays a role of the momentum relaxation time and the longitudinal resistivity can be estimated as 
\be
\rho = m^\star/n_e e^2 \tau_E = (\hbar/e^2) (n_i/n_e) \edc \sim I/B\,.
\label{eq1}
\ee
According to \req{eq1}, $\rho$ is a linear function of $I$ with the slope, $d\rho/dI$, proportional to $1/B$.
This is exactly what we have observed in our experiment, as demonstrated in \rfig{fig4}.
The slope of the linear fit in \rfig{fig4}(b) is given by $2\sqrt{2\pi}\hbar m^\star n_i/e^4 n_e^{3/2} w$, c.f. \req{eq1}, from which we extract $n_i\approx 6 \times 10^7$ cm$^{-2}$. 
This value agrees well with $n_i=n_{\rm 3D} a_B \sim 10^7-10^8$ cm$^{-2}$, estimated using concentration of residual impurities $n_{\rm 3D} \sim 10^{13} - 10^{14}$ cm$^{-3}$ in modern GaAs quantum wells \cite{umansky:2009,umansky:2013,manfra:2014} and $a_B \approx 10$ nm. 

When the current increases such that $\delta > 2\pi a$, the drift will carry an electron away from the impurity after a single collision \cite{baskin:1978}.
Indeed, if $\delta =2 \pi a$, $v_d = \oc a$, and $\tau_E = (2\rc n_i \oc a)^{-1} = (2an_i v_F)^{-1} = \tau$, which is the usual momentum relaxation time at $B=0$.
As a result, at $I > I_c = e^2 n_e B a w/m^\star$, the resistivity is described by the Drude formula and no longer depends on $I$ or $B$.

Taking $a$ as the 2D screening length, $a_B \approx 10$ nm, we find $I_c = \gamma B$, with $\gamma = 0.25$ mA/kG.
This result suggests that, at $B = 0.5$ kG, $\rho(I)$ should saturate at a rather small current $I \sim I_c \approx 0.13$ $m$A.
As no saturation is seen at currents up to 0.35 mA, we can conclude that the typical size of the scatterer in our 2DES is larger than 30 nm, i.e. that these scatterers are distinct from the residual ionized impurities residing in the 2D channel.
We can estimate the typical radius of the scatterer from the zero-field mean free path $l =\tau v_F \approx 9$ $\mu$m.
Using $2 a l n_i = 1$ and $n_i \approx 6\times 10^7$ cm$^{-2}$, we obtain $a \approx 0.1$ $\mu$m, which is an order of magnitude larger than $a_B$.

Having obtained the concentration $n_i$ and the radius $a$ of scatterers from nonlinear transport, we can now look at the CNMR from a different perspective.
As discussed in \rref{shi:2014a}, one of the most peculiar features of the CNMR is that, the characterstic field $B_\star \sim 1$ kG is much smaller than the theoretical prediction considering a combination of smooth disorder, due to remote donors, and sharp disorder, due to unintentional charged impurities (of radius $a_B \approx 10$ nm).
This discrepancy can be reconciled if strong scatterers, with a much larger size $a \gg a_B$, are added into the picture.
At $B = 0$, it is primarily scattering with these large, but dilute, scatterers that is responsible for large resistivity. 
As the magnetic field is turned on, electrons will circle around the scatterers in rosette-like trajectories and the conductivity is reduced.
Above the critical field $B_c = 4.18 \frac{\hbar}{e}\sqrt{n_i n_e}$ \cite{bobylev:1995}, the rosettes no longer form an infinite cluster and the resistivity saturates at $\rho = \rho_\star \ll \rho_0$, whose value is determined only by remote donors and background impurities.
Taking $B_c = B_{\star} \approx 1$ kG, the field where the resistivity minimum occurs, we obtain $n_i\approx 4 \times 10^7$ cm$^{-2}$, in good agreement with the value estimated from nonlinear transport.

The presence of large scatters randomly distributed in the 2DES can explain the absence of HIRO in our 2DES.
Randomness of the distribution could result in spacial variation of the current density $j$, which affects the local electric field $E = jB/n_e e$ and, therefore, $\edc$.
Since $\edc$ enters the cosine in \req{eq.hiro}, HIRO will be strongly suppressed when measured in macroscopic samples.
It appears possible that the weak oscillations shown in \rfig{fig3}, might, in fact, originate from HIRO, modified by nonuniform current distribution.
Since in the presence of large scatterers the average current density can be larger than $I/w$, HIRO extrema should appear at magnetic fields higher than predicted by \req{eq.hiro}, consistent with our observations.
Finally, we note that the reduction of the critical current of the quantum Hall effect breakdown, observed in 2DES with random antidots, has been explained by the inhomogeneous current distribution \cite{nachtwei:1997,nachtwei:1998}.

\begin{figure}[t]
\includegraphics{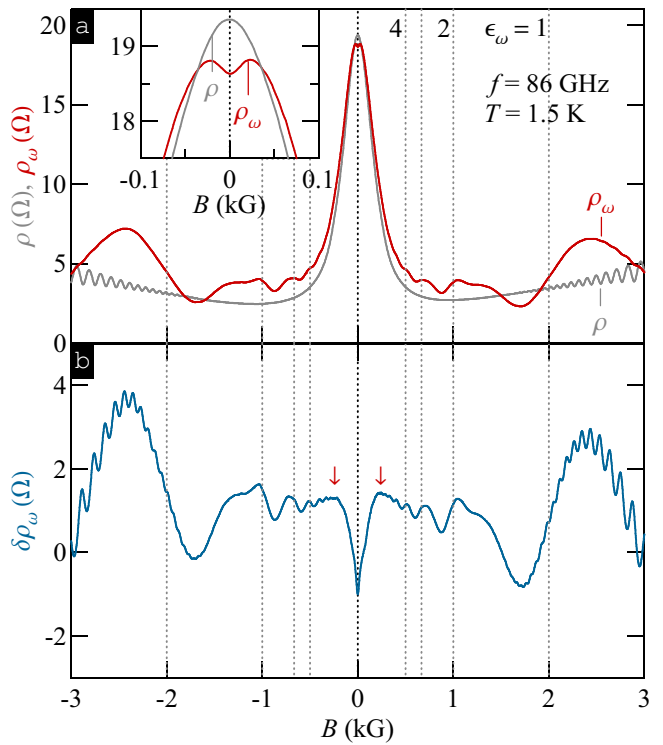}
\vspace{-0.05 in}
\caption{(Color online) 
(a) Resistivity $\rho_\omega(B)$ [$\rho(B)$] measured with [without] with microwave irradiation of frequency $f = 86$ GHz at $T = 1.5$ K. 
Inset is a zoom-in around zero field. 
(b) Photoresistivity $\delta\rho_\omega(B) = \rho_\omega(B) -\rho(B)$. 
Arrows mark the onset of MIRO at $B = 0.28$ kG which corresponds to $\eac \approx 7$.
}
\vspace{-0.15 in}
\label{fig5}
\end{figure}

While no clear signature of HIRO is observed, our 2DES easily reveals microwave-induced resistance oscillations (MIRO) \citep{zudov:2001b,ye:2001}, which occur in the photoresistivity $\delta\rho_\omega = \rho_\omega - \rho$.
Here, $\rho_{\omega}$ and $\rho$ are the resistivities measured with and without microwave radiation, respectively.
The functional form of MIRO is given by $\delta\rho_\omega \sim - \rho_0 \lambda^2 \eac \sin 2\pi\eac$, where $\eac = \omega/\oc$, and $\omega=2\pi f$ is the microwave frequency.
In \rfig{fig5}(a) we present $\rho_{\omega}(B)$ for $f = 86$ GHz and, for comparison, $\rho(B)$.
Direct examination of the data reveal that, for the most part, $\delta\rho_\omega >0$, as illustrated in \rfig{fig5}(b), with a few notable exceptions.
First, we observe $\delta\rho_\omega < 0$ at the first MIRO minimum, close to $\eac = 5/4$, as expected. 
Second, negative photoresponse is observed close to $B = 0$.
Here, microwave radiation reduces the resistivity [cf. inset in \rfig{fig5}(a)], similar to temperature \citep{shi:2014a} and current [cf. \rfig{fig1} and \rfig{fig2}(b)].
The reduction of the resistivity near $B=0$ and its increase almost everywhere except the strongest MIRO minima are consistent with the effects of elevated $T$ on the CNMR \cite{shi:2014a} caused by radiation absorption.
Suppression of Shubnikov-de Hass oscillations in irradiated 2DES gives independent support for electron heating.

Despite a modest mobility of our 2DES, MIRO are observed up to the seventh order, persisting to $B \approx 0.28$ kG. 
Setting $\oc\tq = 1$ at this onset field, we estimate $\tq \approx 14$ ps, a value which is typical for ultra-high mobility 2DES, where HIRO are routinely observed \citep{yang:2002,zhang:2007a,hatke:2009b}.
This result is not surprising since dilute scatterers are not expected to significantly affect $\tq$, lending
further support to our conclusion that HIRO are likely absent in our 2DES because of nonuniform current distribution. 

Finally, we note that our main finding, $\rho \sim I/B$, reminds us of the linear $T$ dependence $\rho(B=B^\star) \sim T$ observed in the same 2DES \cite{shi:2014a}. 
While the role of $I$ can be explained by the $\bf{E} \times \bf{B}$ drift \cite{baskin:1978}, there exist no theories which examine the effects of interactions at finite $T$ on classical memory effects. 
These interactions could, in principle, assist in delocalizing electrons in a nontrivial way, giving rise to observed $T$ dependence (see \rref{shi:2014a} for further discussion).
Indeed, just like the scattering rates due to sharp and smooth disorder cannot be summed up at finite $B$
\cite{mirlin:2001,polyakov:2001}, one cannot simply add electron-phonon and/or electron-electron interactions.

In summary, we have studied nonlinear magnetotransport in a Hall bar-shaped 2DES exhibiting colossal negative magnetoresistance.
We have found that in a wide range of magnetic field, corresponding to the broad resistivity minimum at zero bias, the resistivity is a linear function of direct current with the slope being proportional to the inverse magnetic field.
The data analysis based on the theoretical model considering classical memory effects in the presence of strong, dilute scatterers \citep{baskin:1978,dmitriev:2001}, suggests that the transport in our 2DES is governed by randomly distributed strong scatterers having a typical radius $a \sim 0.1$ $\mu$m and concentration $n_i \sim 10^{8}$ cm$^{-2}$.
This conclusion is further supported by a reasonable value of $B_c$, the absence of HIRO, and complementary photoresistance measurements. 

We thank A. Hatke for device fabrication, Q. Ebner, G. Jones, S. Hannas, P. Martin, T. Murphy, J. Park, and D. Smirnov  for assistance with measurements, and I. Dmitriev, M. Dyakonov, and B. Shklovskii and  for discussions.
The work at Minnesota was supported by the NSF Grant No. DMR-0548014. 
The work at Princeton was partially funded by the Gordon and Betty Moore Foundation and the NSF MRSEC Program through the Princeton Center for Complex Materials (DMR-0819860).
A portion of this work was performed at the National High Magnetic Field Laboratory (NHMFL), which is supported by NSF Cooperative Agreement No. DMR-0654118, by the State of Florida, and by the DOE.
Q.S. acknowledges Allen M. Goldman fellowship.

\end{document}